# Coupled Infrared Imaging and Multiphysics Modeling to Predict Three-Dimensional Thermal Characteristics during Selective Laser Melting

Vijay Kumar[1], Kaitlyn M. Mullin[2], Hyunggon Park[1], Matthew Gerigk[1], Andrew Bresk[1], Tresa M. Pollock[2*], Yangying Zhu[1*]

[1]Department of Mechanical Engineering, University of California Santa Barbara, Santa Barbara & 93106, USA.

[2]Materials Department, University of California Santa Barbara, Santa Barbara & 93106, USA.

*Corresponding author. Email: tresap@ucsb.edu, yangying@ucsb.edu

**Laser heating during additive manufacturing (AM) induces extreme and transient thermal conditions which critically influence the microstructure evolution and mechanical properties of the resulting component. However, accurately resolving these conditions with sufficient spatiotemporal accuracy remains a central challenge. We demonstrate a unique approach that couples high-speed infrared imaging, during selective laser melting of MAR-M247, with a transient three-dimensional (3D) multiphysics simulation to reconstruct the dynamic sub-surface temperature distribution of the melt pool. This integrated framework enables the estimation of experimentally-validated, 3D solidification conditions—including solidification velocities and cooling rates—at the solid-liquid interface while also significantly lowering computational cost. By quantifying solidification conditions, we predict variations in microstructure size and orientation driven by laser processing parameters, and validate them with *ex situ* scanning electron microscopy and electron backscatter diffraction maps. Our findings substantiate that an inte-**

**grated experimental-computational approach is crucial to realize *in situ* prediction and optimization of microstructures in commercial AM.**

Metal additive manufacturing (AM) is industrially relevant for fabricating components with complex geometries that demand reliable mechanical properties (*1, 1–4*). Selective laser melting (SLM), also known as laser powder bed fusion (L-PBF), is one such AM technique that utilizes a moving high power laser beam, which melts and fuses the metal to form a desired component geometry on a layer-by-layer basis (*5, 6*). The high power laser spot rapidly creates a localized temperature field, with temperatures exceeding the melting point. During SLM, local thermal gradients and solidification velocities at the solid-liquid (S-L) interface play a crucial role in the microstructural evolution and occurrence of defects in printed components (*4, 7–14*). Therefore, obtaining the transient three dimensional (3D) temperature distribution during solidification is crucial to control such properties and mitigate defects.

The surface temperatures within the melt pool are typically measured using non-invasive techniques such as near/mid infrared (IR) thermography (*15–21*), two-wavelength thermometry (*10, 22, 23*), and pyrometers (*24, 25*), all of which face challenges with sufficient spatial and temporal resolution for the measurement of solidification conditions. Pyrometers are often limited in spatial resolution, with the spot size on the order of melt pool width (approximately 100 µm), providing a spatially averaged temperature and thereby restricting the estimation of accurate thermal gradients (*25*). The use of two pyrometers has allowed for gradient detection but has been limited to two-point detection with one at melt pool center and other at the tail. (*24*) Recently, a color camera based two wavelength (red and green) method has been demonstrated to measure the melt pool peak temperature without the need for emissivity calibration, though is accompanied with large uncertainties between 300 - 400 K. The peak spectral emissive radiation at a temperature of 1700 K, which is close to the melting point of numerous AM materials, is at 1.7 µm in the infrared region. Therefore, red and green color based two wavelength method suffers significant background noise near or below the melting and solidification temperature, which is crucial for understanding solidification conditions. Additionally, studies in the literature have attempted to utilize IR imaging to capture the transient melt pool surface radiation to estimate the melt pool surface temperature (*16, 26*) and solidification conditions (*16*). While informative, these studies are often limited to low spatio-temporal resolution, insufficient for accurate determination of thermal gradients (*16, 17*). Recently, Gould et al. implemented a coupled high spatio-temporal IR imaging (> 10000 fps and 30 μm) and X-ray imaging during SLM (*26*), which enabled the determination solidification conditions at the melt pool surface with fine resolution.

Further, IR imaging requires careful calibration to convert the measured radiation signal to the true temperature, which is often non-trivial for high melting-temperature alloys that often lack carefully calibrated emissivity values. The Stefan-Boltzmann law, which has been utilized for previous investigations (*17, 18, 26*), is applicable only for situations where the acquired radiation contains the entire electromagnetic spectrum. However, most commercially available IR imaging systems can acquire only a range electromagnetic spectrum such as: Near IR, Short-wavelength IR, Mid-wavelength IR, or Long-wavelength IR. Such conditions necessitate the use of Planck's law, integrated only for the range of wavelength being acquired by the IR imaging system. Very recently, X-ray radiography has been shown to measure the melt pool sub-surface temperature along the melt pool cross-section (*27*). Critically, information provided by the sophisticated technique and instrumentation is limited to a singular cross section plane.

In addition to the temperature distribution, the solid-liquid (S-L) interface velocity, or the solidification velocity, also influences microstructural evolution (*28*). Recently, X-ray radiography has been used to monitor the melt pool evolution (*6*), sub-surface S-L interface velocity (*29*) and fluid flow inside the melt pool using tracer particles (*30*). As mentioned, X-ray radiography provides single plane information and is prone to low contrast between solid and liquid regions, adding uncertainty to analysis (*9, 29*). Coaxial cameras have also been shown to visualize and monitor the melt pool top surface area (*31*) and deposition layer height (*32*) to control the laser power during direct energy deposition (DED) (*31, 32*) and laser powder bed fusion (LPBF) (*33*). In summary, these investigations have been able to visualize and evaluate the solidification velocity, but are limited to a single plane, either on the melt surface or a singular cross section.

While there has been significant progress of measuring surface temperatures during melting, these approaches do not fully encompass the three-dimensional, transient nature of solidification. Due to experimental difficulty in obtaining the 3D solidification conditions at the S-L interface, high-fidelity numerical simulations (*5, 28, 34–37*) have been the method of choice to compute spatio-temporal thermal gradients and solidification rates. The numerical simulations involve absorption and powder interactions (*38, 39*) coupled with thermofluidic transport (*5, 34, 40*) in the melt pool. Numerical simulations have also been used to understand the mechanisms governing defect formation, such as keyholing (*13*), balling (*36*), and lack of fusion porosity (*36*). While valuable, pure numerical simulations can be time consuming, computationally expensive, and may be too intensive to enable real time control of processing parameters. Additionally, due to the lack of known thermophysical properties that are involved with conduction, convection, radiation, evaporation, and fluid flow (*41*), it can be challenging to mimic experimental conditions, and often requires artificial parameter adjustments to match experiments.

In summary, state of the art experimental techniques generally provide two dimensional (2D, either surface or cross-section) information to evaluate the dynamics of AM process, often accompanied by high degrees of uncertainty. Numerical simulations, alternatively, provide 3D information at the expense of time and computational cost, and may be subject to inaccuracy. To address this gap, we develop and validate a unique approach that combines 2D transient melt pool surface radiation maps, obtained using high-speed IR thermography, with 3D heat transfer modeling to unveil spatio-temporally resolved solidification conditions (i.e., thermal gradients and solidification velocity) at the S-L interface. In this work, we perform single track SLM experiments on a wrought nickel-based superalloy MAR-M247 substrate, which is an excellent material for high temperature components in aerospace applications but is susceptible to crack defect formation during printing (*42–44*). To obtain an accurate temperature field of the melt pool surface, we integrate Planck's law with additional considerations, like in-band irradiance, non-ideality of optics, and emissivity. Next, we import the experimentally obtained time-resolved surface temperature into a 3D multiphysics model to predict the melt pool 3D temperature distribution, S-L interface profiles, temperature gradients, and solidification velocities during SLM process of MAR-M247 alloy, which are coupled with microstructural analyses to inform predictive microstructural mapping. The technique presented herein provides multi-dimensional data-rich outcomes that can be used as a platform to readily develop real time defect detection platforms and enable guided microstructure predictions.

## Results

A schematic of the melt pool surface radiation intensity map acquired during experiments, and the resulting 3D melt pool profile estimated from the 2D surface IR imaging, is shown in Figure 1a (experimental setup in Figure S1). Briefly, first, to convert the radiation energy acquired by the IR camera to surface temperature, we create a custom calibration by combining Planck's law and IR camera response to account for the optics transmission (Telops Microscope 1X lens) and sensor exposure time (15 µs) used in the experiments. The specific details of melt pool surface radiance ($W/m^2.sr$) conversion to temperature maps using Planck's law (integrated between 3 - 5.5 µm) for the current work, a comparison with Stefan-Boltzmann-based conversion is provided in Supporting Information Note 2 (Figure S2, S3, and Table S1). A generalized methodology and detailed uncertainty analysis for the same can be found in our previous work (*45*).

We then import the 2D temporally varying (15,000 fps) melt pool surface temperature to a COMSOL heat conduction and solid-liquid phase change model to generate the time-dependent sub-surface 3D temperature distribution. Once solved at each time step (same as the IR camera time step, 0.06 ms), the solidus temperature (1573 K for MAR-M247) contour is then extracted as the 3D melt pool boundary (3D melt pool profile shown in Figure 1b). A constant solidus temperature is used to identify the S-L interface. Throughout this work, the laser scanning is taken as X direction, melt pool width as Y direction, and melt pool depth as Z direction, and are used interchangeably in the text.

Using this methodology, a comparison of the melt pool cross-section profile for a laser processing condition of 200 W and 500 mm/s, obtained from COMSOL model and ex situ SEM, is presented in Figure 1c (Supplementary video 1). Evidently, the accuracy of the spatio-temporally resolved melt pool dimensions is parallel to those achieved by high fidelity modeling, despite employing less input parameters and computational power. Temperature independent thermophysical properties were employed to lower the computational complexity associated with the simulation. To analyze the sensitivity of the approach to variations in thermophysical properties, the model was evaluated with a sweep of input thermophysical properties as shown in Figure 1d. Critically, the melt pool depth variation is within 1.5 µm across a relatively large variation (10% variation shaded in gray) of thermophysical property inputs. This low variation is especially important for deter-mining the solidification conditions of materials lacking measured thermophysical properties, like many high temperature metals (*46*). The details of the coupled COMSOL model, thermophysical properties used, and impact of latent heat incorporation on the model are presented in Supporting Information Note 3 (Figure S4, S5, and Table S2). A direct comparison of melt pool depth (Figure 1e) obtained from COMSOL and ex situ SEM for various laser processing parameters highlights the similarity of the outputs from modeling to experimental measurements. This is supplemented by a comparison of melt pool cross section profiles obtained from COMSOL and ex situ SEM for six different laser processing parameters (Figure S6). The coupled COMSOL modeling approach works well for a multitude of experimental conditions because the dynamically changing melt pool surface temperature, which is the surface boundary condition, directly captures the influence of laser absorption, surface to ambient radiation, natural convection, and thermal Marangoni. While the high fidelity models in the literature have relied on assumptions to incorporate the impact of these challenging-to-obtain parameters, the current work exploits experimental measurements to forego the related assumptions in numerical simulations.

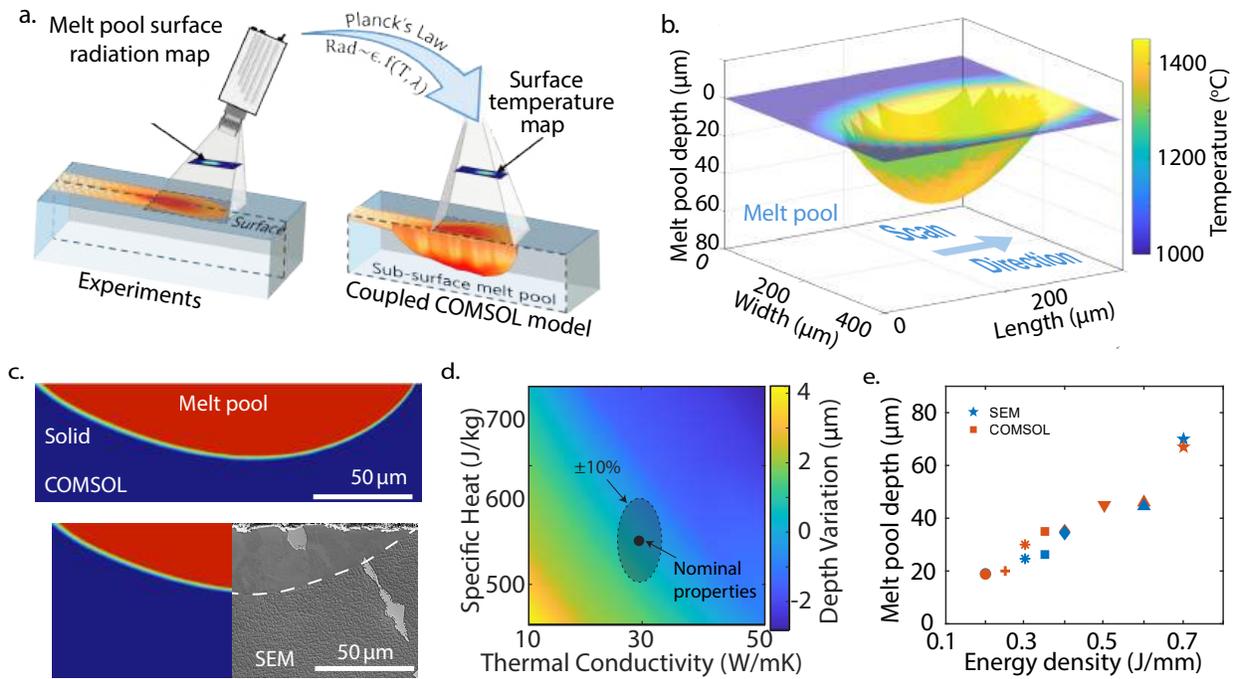

**Figure 1**: **A coupled experimental-modeling approach to extract the dynamic 3D melt pool profile and temperature distribution during laser melting.** a. Illustration of melt pool surface radiation conversion to sub-surface (3D) melt pool profile. b. 3D melt pool profile (S-L interface) obtained using couple COMSOL model for 200 W and 500 mm/s. c. Comparison of melt pool cross-section profile obtained using our coupled COMSOL model and ex-situ SEM, for 200 W and 500 mm/s. d. Parametric sweep showing sensitivity of melt pool dimensions to a 10% variation in thermophysical properties of Mar-M247. e. Comparison of melt pool depths extracted from coupled COMSOL model in comparison ex-situ cross-section SEM measurements. The shape of markers represent individual laser processing condition as energy density (laser power/laser scan speed).

Thermal Marangoni flow, promoted by surface tension gradients in the liquid, is expected to influence the fluid flow and thermal conduction within the melt pool (*5, 40, 40, 47–50*), but ascertaining the magnitude of Marangoni flow has remained challenging. Here, we hypothesize that the IR surface measurements, used as the transient top surface boundary condition, partially account for changes to heat transfer by Marangoni flow. To substantiate the hypothesis, we develop a coupled COMSOL model with additional thermal Marangoni flow inside the melt pool. The temperature distribution inside the melt pool (200 W and 500 mm/s) with and without the thermal Marangoni flow is shown in Figure 2a and 2b. The surface tension coefficient for thermal Marangoni, in Figure 2b, is taken as 0.01 (mN/mK) for a Ni superalloy from the literature (*40*). This also aligns with the measured decrease in surface tension with increasing surface temperature for a similar Ni superalloy, Inconel 718 (*51*). The temperature distribution inside the melt pool is slightly more uniform (Figure 2b) with thermal Marangoni flow, which is consistent with convective heat transfer induced by Marangoni flow. A negative surface tension gradient results in the fluid flow from melt pool center towards the edge of the melt pool, forming two circulation zones inside the melt pool as shown by arrows in Figure 2c (velocity magnitude in Figure S7).

Convective flow can become a more dominant heat transfer mechanism where the magnitude of Marangoni flow is largest. Using spatial analysis of the Peclet number, a measure of the relative strength of convective to conduction heat transfer, the influence of convection is spatially analyzed throughout the melt pool in Figure 2c. The Peclet number is computed as $Pe = L.V/\alpha$, where $L$ is characteristic length, $V$ is characteristic flow velocity, and $\alpha$ is thermal diffusivity of liquid. Here, the Peclet number is predominantly influenced by changes in fluid flow velocity. The high fluid flow velocity, $V_{\max}$= 1.1m/s (Figure S7), and thus Peclet number, is limited to the melt pool top surface $Pe_{\max}$=26 (Figure 2c). In contrast, the average fluid flow velocity inside the bulk of the melt pool is $V_{avg} = 0.043$ m/s, with a corresponding Peclet number $Pe_{avg}$ of 0.5. A $Pe_{avg} < 1$ indicates that for the conditions tested, on average, conduction is still dominant inside the melt pool, except near the top surface. Therefore, measuring surface temperature with sufficient temporal resolution can effectively capture, to great extent, the impact of melt pool fluid flow and convective effects. This is quantified in Figure 2d wherein the cross-sectional profiles in the model, with and without additional Marangoni flow modeling incorporated, are nearly identical. Additionally, the width of the melt pool remains the same between both modeling approaches due to the experimental top surface temperature defined as the top surface boundary condition. Given that the current work focuses on the thermal gradients and solidification velocity at the S-L interface, which remains similar for both cases, we employ the conduction and phase change model (thermal Marangoni model being > 70× time consuming).

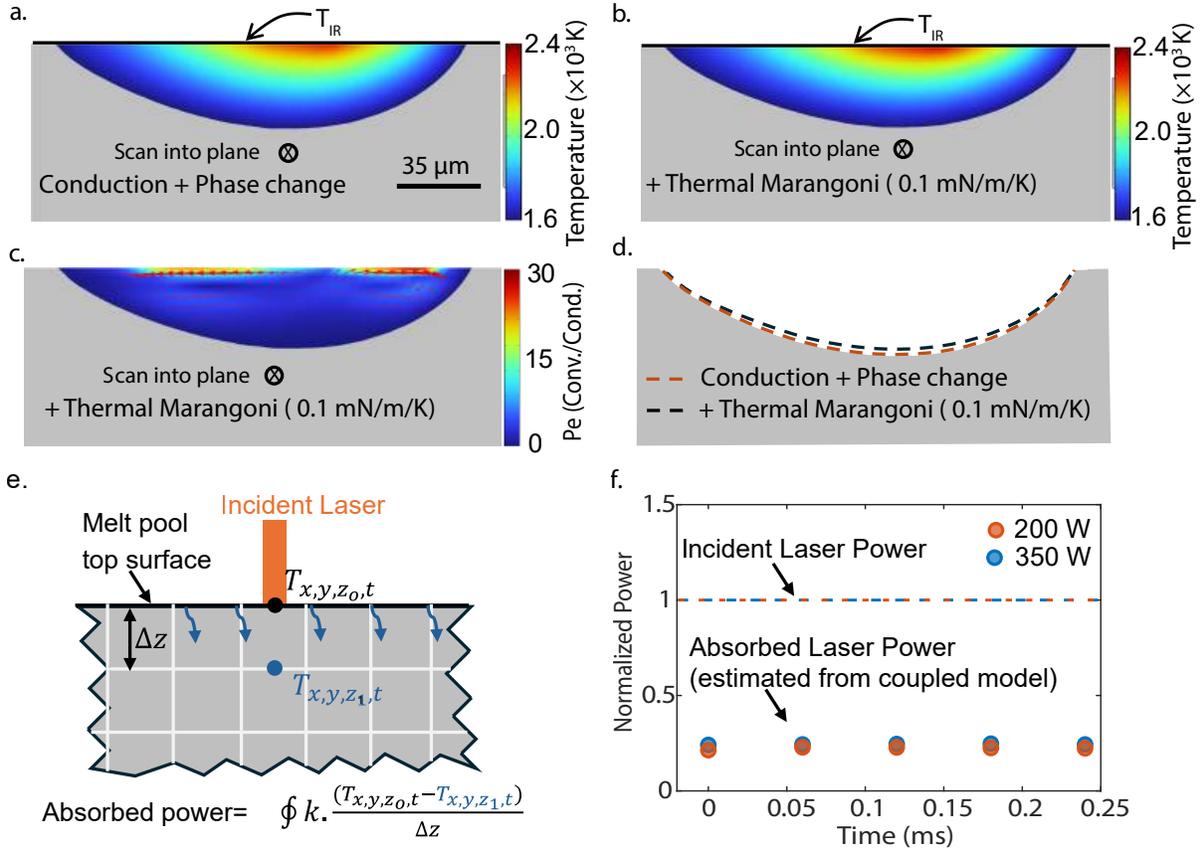

**Figure 2**: **Experimental surface temperature, imported as boundary condition to simulations, captures energy balance and fluid flow at the melt pool surface.** Temperature distribution inside the melt pool, a. obtained from a conduction and phase change model and, b. incorporating thermal Marangoni for 200 W laser power and 500 mm/s laser scan speed. The temperature distribution with thermal Marangoni is slightly more uniform. c. The magnitude of *Pe* along with the direction of the in-plane fluid flow which results in the redistribution of the temperature inside the melt pool. The highest convective flow, characterized by high *Pe*, is localized only to the melt pool surface. d. Comparison of melt pool cross section profile with and without thermal Marangoni. The melt pool dimensions and thermal profile remain largely similar with the addition of thermal Marangoni to the conduction only model. e. Schematic to represent absorbed laser energy estimation from coupled model. f. Absorbed laser power estimated from coupled model for two extreme power cases in this work.

The coupled COMSOL model incorporates the laser absorption and effective surface emissivity. To obtain the absorbed laser power, we integrate the z-direction surface heat flux over the melt pool surface area (Figure 2e). The proof-of-concept laser energy deposition estimation shows that for a 200 W incident laser power only ≈ 50 W is absorbed on the surface (Figure 2f) whereas for 350 W incident laser power ≈ 80 W (Figure 2f) is absorbed on the melt pool surface. Interestingly, absorbed laser power is close to the typical absorptivity assumption of ≈ 0.3 used in the literature (*35, 40*).

The direct coupling of experimental temperature measurements on the surface with relatively simple computational modeling yields accurate melt pool geometries and insights to absorption and convective heat transfer without the use of artificial parameter adjustments. The ability to capture the melt pool dimensions, temperature distribution, and fluid flow in three dimensions with high accuracy and temporal fidelity enables the direct calculation of solidification velocity (***R***), which describes the speed and direction of the S-L interface motion during solidification. The solidification velocity (Equation 1) can be obtained by equating the total derivative of the interface temperature ($T_{x_i, y_i, z_i, t}$) to zero as the interface temperature remains constant in space and time.

$$\mathbf{R} = -\frac{\frac{\partial T(x_i, y_i, z_i, t)}{\partial t}}{|\nabla \mathbf{T}|} (\hat{\mathbf{n}}_\mathbf{T}) \qquad (1)$$

where $\partial T/\partial t$ is the cooling rate, $|\nabla \mathbf{T}|$ is the magnitude of total thermal gradient in 3D ($\sqrt{G_x^2 + G_y^2 + G_z^2}$), $\hat{\mathbf{n}}_\mathbf{T}$ is the unit thermal gradient vector, and $G_x$, $G_y$, $G_z$ are thermal gradients in X, Y, and Z direction. During solidification, the direction of solidification velocity is inferred by the direction of maximum thermal gradient ($\hat{\mathbf{n}}_\mathbf{T}$ in Equation 1). Details for the estimation of $\mathbf{R}$, $\nabla \mathbf{T}$, and $\partial T/\partial t$ at the S-L interface are in Supporting Information Note 6 (Figure S8).

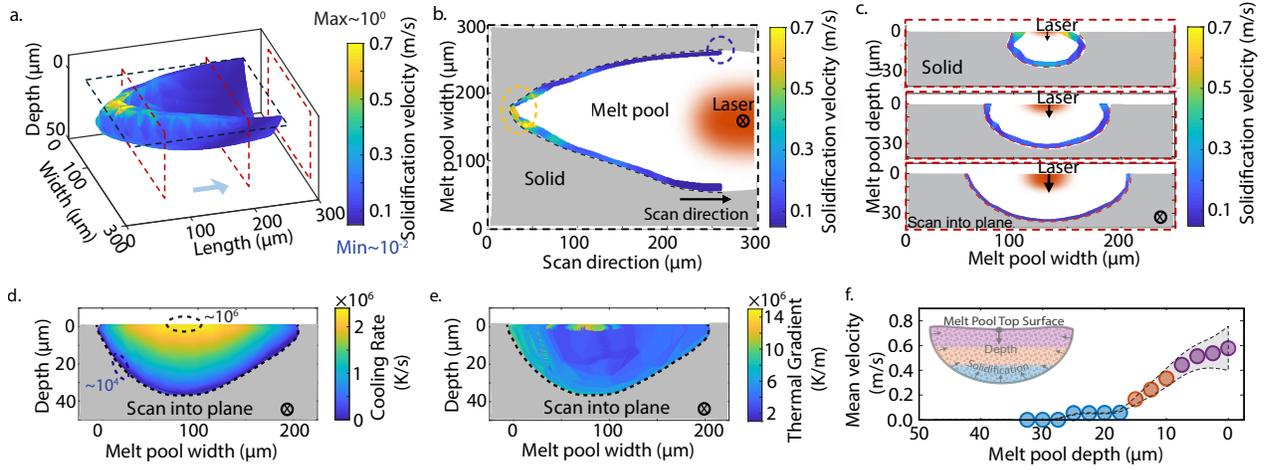

**Figure 3**: **Solidification velocity estimation using coupled model.** a. 3D solidification velocity along the melt pool boundary. The experimental conditions are 200W laser power and 500 mm/s laser scan speed. b. Solidification velocity variation along the melt pool top surface (XY view). c. Solidification velocity along the cross-sectional boundary of the melt pool (YZ view). The red dashed curves represent the cross-sectional S-L boundary at various time instants, as the melt pool shrinks. d. Cumulative representation of temporal evolution of cooling rate at the S-L interface as solidification progresses. e. 3D thermal gradients at the melt pool boundary during various stages of solidification. f. Mean solidification velocity along the depth of the melt pool. The standard deviation of the solidification velocity increases as we move towards the surface, highlighting increasing non-uniformity in solidification process.

The variation in the magnitude of $\mathbf{R}$ along the melt pool boundary for a laser power of 200 W and laser scan speed of 500 mm/s is shown in Figure 3a. Specifically, as shown in the top down view of the melt pool top surface boundary (XY plane) in Figure 3b, $\mathbf{R}$ changes significantly by 3 orders of magnitude from $10^{-3}$ m/s, at the edge of the melt pool (blue dashed circle in Figure 3b), to $\sim 10^0$ m/s at the tail of the melt pool (yellow dashed circle in Figure 3b). As the laser moves forward, the sub-surface (YZ plane) melt pool boundary shrinks inwards and towards the top surface as shown by the red dashed curve in Figure 3c. At any one particular red dashed curve in Figure 3c, representing the S-L interface at a given time, a larger solidification velocity is observed at the surface edges in comparison to the bottom of the melt pool. This increase in **R** is primarily driven by a substantial increase in the cooling rate ($dT/dt$ in Figure 3d changes from $\sim 10^4$ K/s to $\sim 10^6$ K/s) from edge to the tail of the melt pool. At the same time, the thermal gradient (Figure 3e) generally decreases as solidification progresses. Further, the mean |**R**| monotonically increases from bottom of the melt pool towards top surface, as quantified in Figure 3f. The mean solidification increases from zero at the bottom of the melt pool to 0.6 m/s at the top of the melt pool, which may promote microstructural transitions near the surface. The standard deviation of the solidification velocity (grey shaded region in Figure 3f) increases markedly. Such an increase in standard deviation means the |**R**| has a larger spread towards the end of solidification, potentially inducing heterogeneity in the microstructure shape and size distribution.

The highly transient motion of the solidification front can be evaluated by investigating the normal direction ($\hat{n}_T$) of the S-L interface. The directional motion of the S-L interface for the melt pool top surface boundary in Figure 3a is given by surface normal arrows in Figure 4a. The solidus isotherms are presented at different time steps during melt pool solidification in Figure 4b. Interestingly, the normal direction of the interface is not always oriented in the scan direction, specifically at the tail of the melt pool. Additionally, the maximum velocity near the tail of the melt pool is greater than the laser scan speed. These observations are somewhat different from the alternative geometrical relation for local velocity, $R = V_{laser} \cos\theta$ where $\theta = G_x/\sqrt{G_x^2 + G_y^2 + G_z^2}$ and $R_{max} = V_{laser}$. In the absence of experimental ground truth, this relation is commonly employed to approximate $R$ using laser scan speed $V_{laser}$ along the solidification interface (*7, 52–54*). It is important to note that the geometrical relation provides a projection of laser scan speed in the solidification direction (direction being $\cos\theta$) and neglects the magnitude of local thermal gradients, a driving force for solidification. The geometrical relation is inherently limited to $R_{\max} = V_{laser}$, however during the experiments we observe $R_{\max}$ up to $\sim 1.5 \times V_{laser}$. These high velocities are observed near the surface, where convection is the more dominant mechanism governing heat transfer. Note that while the $R_{\max}$ from our coupled modeling is greater than the laser scan speed, the median solidification velocity ($\sim$110 mm/s) remains lower than the laser scan speed (500 mm/s).

Since the coupled experimental and modeling approach yields three dimensional temperature distributions with fine spatial and temporal resolution, local changes in thermal gradients and velocities, and therefore in microstructure, can be qualitatively evaluated. One important aspect of microstructural formation during solidification, and AM more generally, is the growth of sub-grain cells and dendrites in the melt pool. The orientation of dendrites and cells is dependent on the 3D thermal conditions and crystallographic considerations. Here, the influence of thermal conditions on the orientations and spacings of solidification cells in a laser melted single crystalline substrate is evaluated. First, the orientations of the grains and cells present within an experimental melt pool cross-section(Figure 4c) are analyzed for 200 W and 500 mm/s. The EBSD pattern shows that the melt pool maintains epitaxial growth with the single crystal substrate orientation (Figure 4d), indicating that changes in the cell directions are predominantly attributed to thermal gradients. It is worth noting that the grain orientation is preserved, even along the boundaries of the larger carbide in the melt pool (Figure 4d), indicating that the carbides do not promote grain nucleation here. It is presumed that solidification cells are present in this melt track, since no secondary arms characteristic to dendrites are observed. The spatial variations in cell growth observed in the experimental cross-section were estimated by measuring the angle between the centerline (Z-direction) and the longitudinal axis of the solidification cell, and are presented in Figure 4e. Along the centerline of the melt pool, cell orientations are largely aligned with the Z-direction. Larger angles are observed near the edges of the boundary, where the direction of maximum thermal gradient exceeds 45 degrees. Some variations in cell orientation are observed immediately surrounding the large carbides.

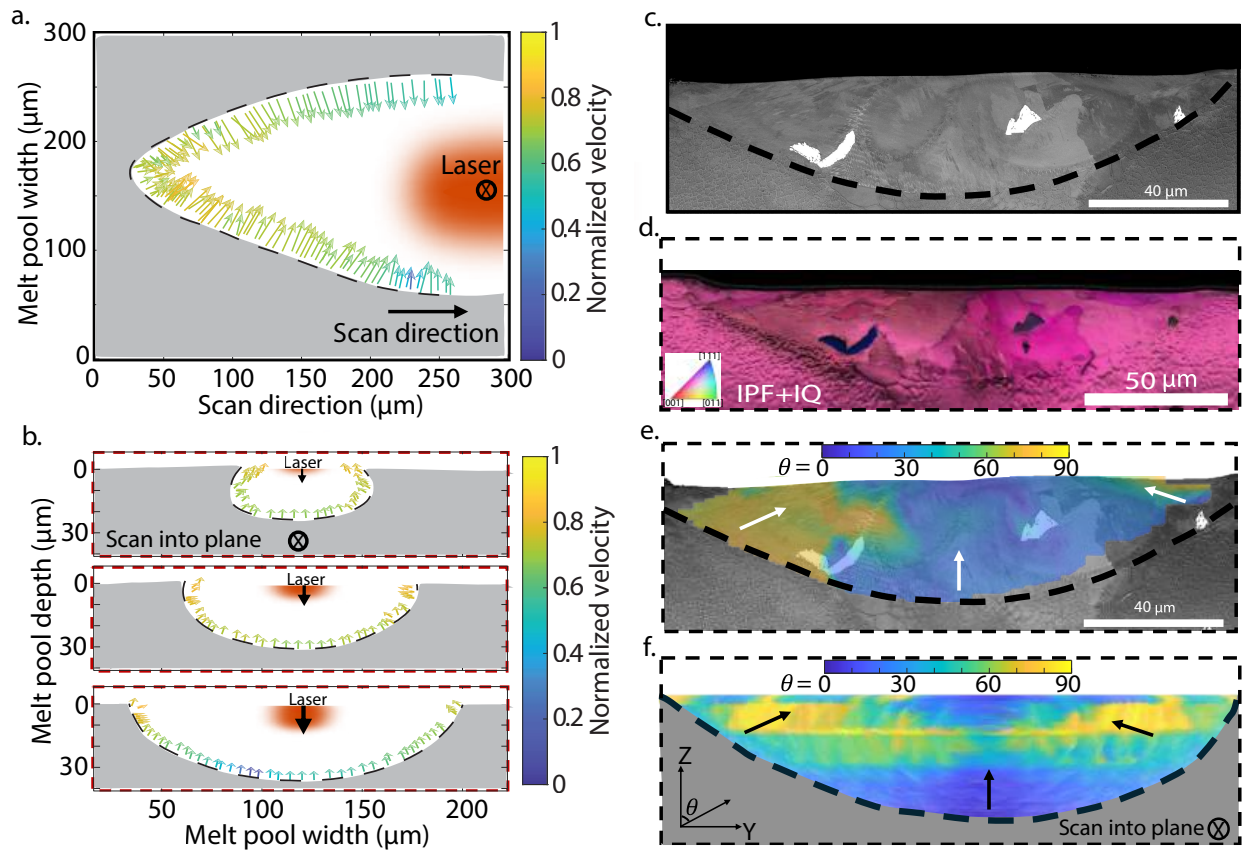

**Figure 4**: **S-L interface direction of motion during solidification.** Solidification velocity vectors direction along: a. the melt pool top surface boundary (top down view), b. the subsurface (cross section) melt pool boundary at various stages of solidification. Each black dashed curve represents the S-L interface at a particular time instant during solidification. The laser power is 200 W and laser scan speed is 500 mm/s. The color bars show solidification velocity values normalized with the maximum solidification velocity. c. Backscatter electron (BSE) micrograph of the melt pool cross-section. d. Ex situ EBSD inverse pole figure (IPF) and image quality (IQ) maps. The relatively same crystallographic orientation across the melt pool suggests that the direction of cell growth is primarily influenced by thermal conditions. e. Cell orientation ($\theta$) with respect to z-axis, estimated from SEM, overlaid on the cross-section. f. Temperature gradient direction with respect to z-axis, estimated from COMSOL model. The temperature gradient direction represents favorable growth direction.

To compare the predicted orientations from the coupled simulations to those in the ex situ cross-section, we compute the favorable growth direction in the cross-sectional plane of the melt pool using Equation 2:

$$\theta = \cos^{-1}(G_z/\sqrt{G_x^2 + G_y^2 + G_z^2}) \qquad 2$$

where $\theta$ is the cell orientation with respect to Z axis. The direction of maximum thermal gradient estimated using COMSOL is illustrated in Figure 4f. At the base of the melt pool, the solidification direction predicts cell growth ($\theta$ in Figure 4f) to be aligned with the z-axis ($\theta = 0°$), consistent with the cellular orientation in Figure 4e. The cell orientation near the edges increases from $0°$ to $\sim 80°$. This is similar to the general trend of increasing growth angle in the SEM, as we move away from the bottom of the melt pool towards the edges. The transition in orientation occurs at similar positions along the melt pool boundary in the experimental and computed cross-sections. In summary, suitable agreement is demonstrated between the experimentally observed and computed orientations for cell growth.

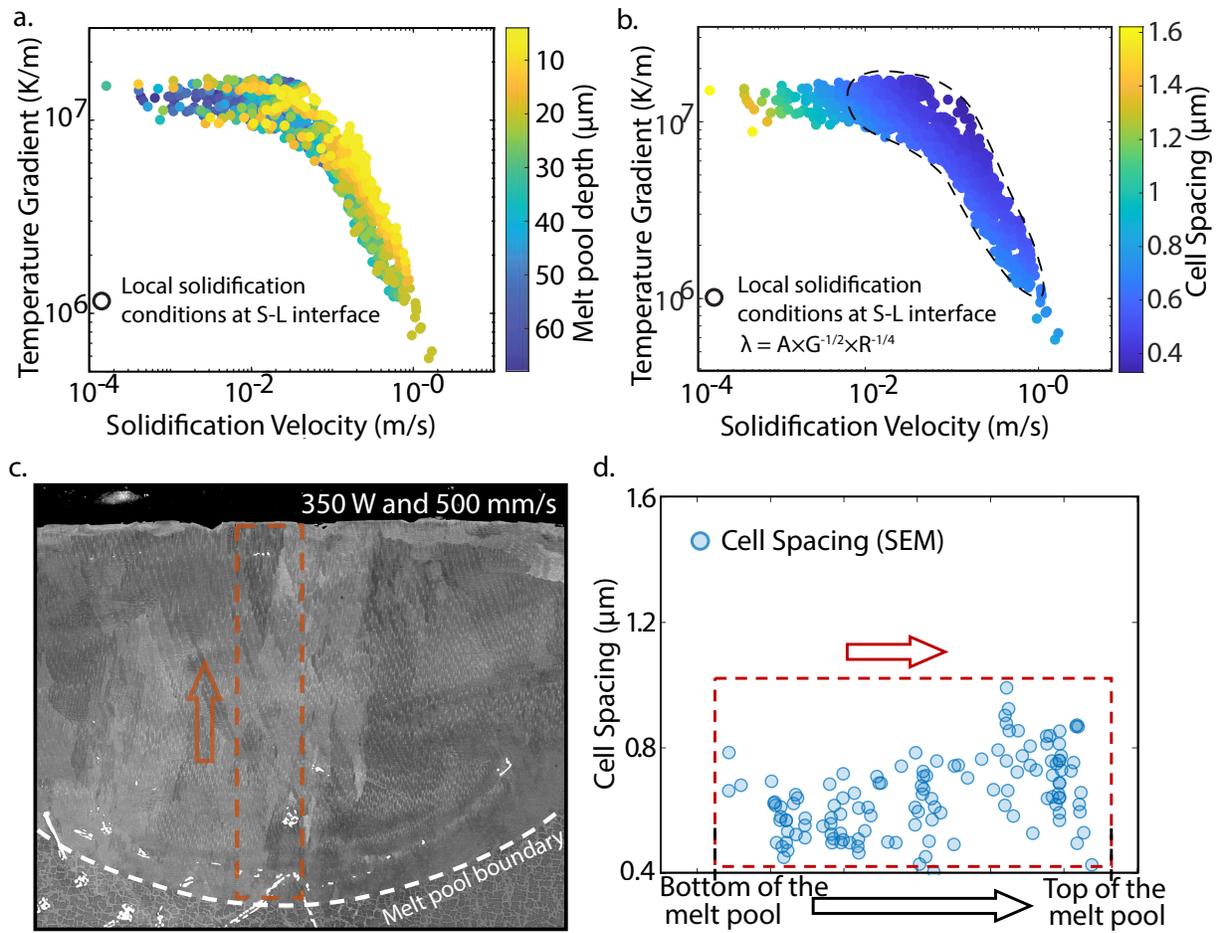

**Figure 5**: **Correlating solidification conditions with ex situ metallography.** a. The variation of solidification thermal gradients and solidification velocity along the melt pool depth. b. Cell spacing (color bar in μm from Trivedi model) overlaid on the solidification conditions data for 350 W and 500 mm/s. The black dashed outline encloses lower 90 percentile of cell spacing c. SEM of a cross section for 350 W and 500 mm/s laser scanning parameters. d. Cell spacing variation along the melt pool depth, obtained from the ex situ SEM (orange dashed region in part c). The position of individual markers is overlayed on the SEM in Figure S10.

Next we map individual solidification conditions $|\nabla \mathbf{T}|$ and $|\mathbf{R}|$ along the melt pool depth (Figure 5a) and combine with Trivedi model (*55*) to compute the spatial variations in cell spacing in the melt pool (Figure 5b). Briefly, the Trivedi model is based on the relationship between dendrite tip radius, growth velocity, and thermal gradient, accounting for solute and thermal diffusion effects during solidification. It allows the prediction of microstructural features like cell spacing using parameters such as the temperature gradient and solidification velocity. A simplified form of Trivedi model relating the cell spacing with thermal gradients and solidification velocity is shown in Equation 4 (*55, 56*):

$$\lambda \text{ (µm)} = A * G^{-0.5} * R^{-0.25} \qquad 4$$

where $\lambda$ is the cell spacing and $A$ is the material dependent constant. The mean $|\nabla \mathbf{T}|/|\mathbf{R}|$ ratio of $\sim 10^8$ obtained from the coupled simulations is lower than the critical constitutive supercooling criteria of $\sim 10^{11}$ (Figure S13), suggesting non planar growth, in agreement with the absence of a planar interface at the base of the melt pool in experiments. We then empirically obtain the constant $A = 700$ K$^{-0.5}$ m$^{0.25}$ s$^{0.25}$ (from Equation 4), such that the mean cell spacing obtained from the coupled COMSOL model (Figure 5b) agrees with the mean cell spacing from ex-situ SEM (Figure 5c and d). Interestingly, for the cell spacing predictions shown in Figure 5b, a constant $A = 700$ K$^{-0.5}$ m$^{0.25}$ s$^{0.25}$ would results in a lower 90 percentile cell spacing ranging between 0.42 µm - 0.67 µm (dash enclosed region in Figure 5b) which is in remarkable agreement with the ex situ SEM image (Figure 5d, EBSD IPF and IPF+IQ maps in S14). This becomes particularly noteworthy as the constant A is based on the mean cell spacing and does not account for the cell spacing distribution range in itself. The lower 90 percentile of the cell spacing in the orange dashed box in Figure 5c SEM image lies between 0.38 µm - 0.81 µm. A comparison of both Figure 5a and b shows that, despite significant change of $|\nabla \mathbf{T}|$ and $|\mathbf{R}|$ as solidification progresses, the predicted cell spacing shows very minor changes with melt pool depth. This outcome is validated by the ex situ SEM (Figure 5c) wherein there is no clear trend of cell spacing variation along the melt pool depth(Figure 5d). The ex situ SEM also reveals a subtle increased spread in the cell spacing range towards the top of the melt pool, which mimics the variability in the solidification velocity shown in Figure 3c.

    To quantify the interrelated differences in laser parameters, thermal conditions, and cell spacings, the solidification conditions for two representative melt tracks (200 W laser power with 500 mm/s and 1000 mm/s scan speed) are illustrated in Figure 6a. For a particular laser track, each marker in Figure 6a represents the $|\mathbf{R}|$ and $|\nabla \mathbf{T}|$ of a local point on the S-L interface. Figure 6a

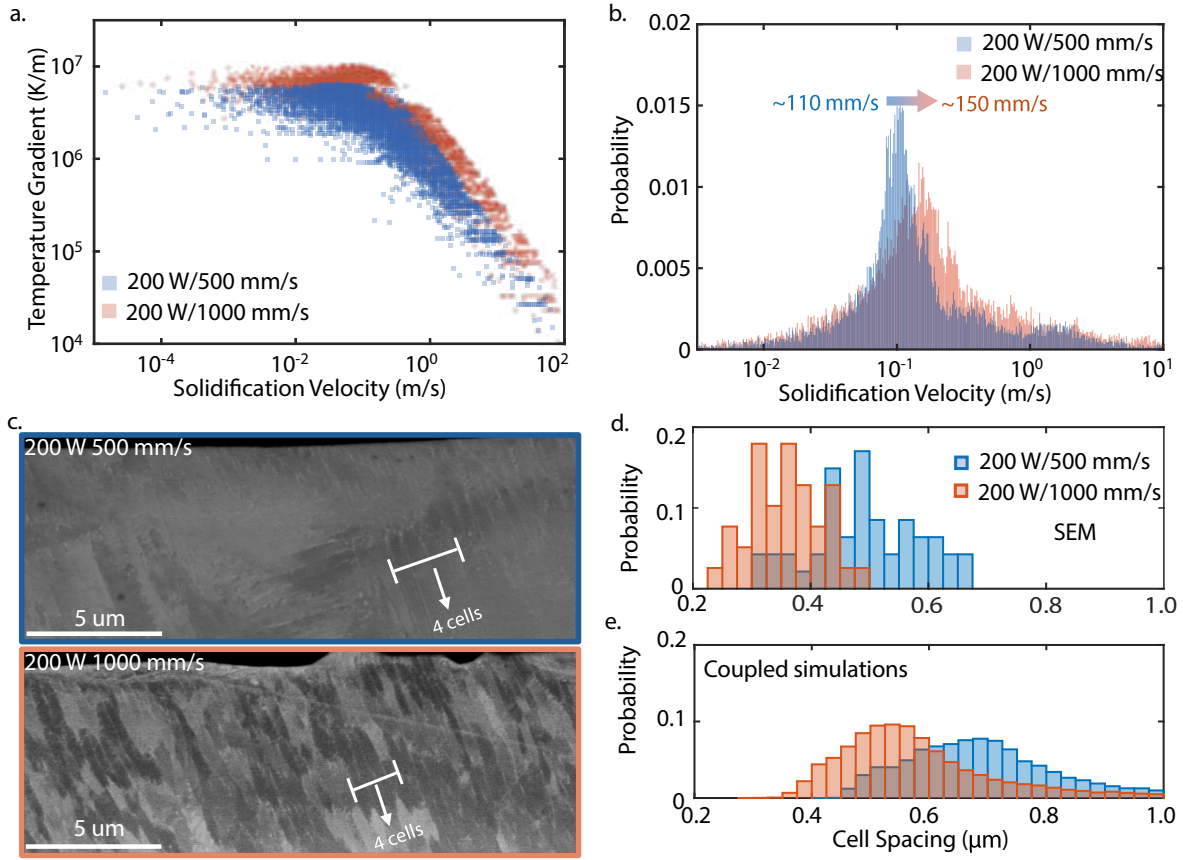

**Figure 6**: **Influence of process parameters on local solidification conditions at the S-L interface.** a. Solidification conditions, at the S-L interface, for laser power of 200 W with scan speeds of 500 mm/s (blue markers) and 1000 mm/s (orange markers). b. Shift in the median |**R**| with increase in laser scan speed. Blue markers represent 500 mm/s and orange markers represent 1000 mm/s laser scan speed. c. SEM of the melt pool cross-section for 200 W laser power with 500 mm/s (blue box) and 1000 mm/s (orange box) laser scan speed. An increase in |**R**| (in part b) corresponds to a reduced cell spacing, highlighted by white lines. Cell spacing distribution comparison of ex situ SEM based estimation in part d with COMSOL prediction in part e. For both d and e, the slower |**R**| for 500 mm/s laser scan speed result in increased cell spacing (blue bars).

shows a collection of |**R**| and |∇**T**| values for all the points on a S-L surface. For a given power, the thermal gradients |∇**T**| increase with the laser scan speed. The dependence of solidification velocities **R**, on laser scan speed, is clarified by the distributions presented in Figure 6b. Here, the median solidification velocity clearly increases from ∼ 110 mm/s to ∼ 150 mm/s with increase in

laser scan speed. Notably, the tail ends of each distribution are similar. While the lower velocity tail of each distribution extends to identical lower velocities, the melt track performed at a higher scan speed (1000 mm/s) extends to higher solidification velocities. The solidification velocity trend is also observed by the spread in Figure 6a however, globally lower and overlapping thermal gradients for 500 mm/s mask the trend.

The difference in cell spacings between the two melt tracks corroborate the respective differences in thermal conditions. In agreement with the higher thermal gradients and solidification velocities measured for the 1000 mm/s melt track, the solidification cells have visually finer spacings in the micrograph (Figure 6c). This is quantified by the measured distribution of cell spacing values, along the centerline for each melt track, presented in Figure 6d (IPF and individual cell spacing in Figure S9).

The utility of the measured thermal conditions to predict microstructure formation is demonstrated in Figure 6e. For 200 W laser power and 500 mm/s laser scan speed, combining equation 4 and data from Figure 6a (using the material dependent constant $A$ from Figure 5), the coupled model predicts a median cell spacing of $\sim$ 0.7 µm (Figure 6e) with $25^{th}$ to $75^{th}$ quartiles between 0.60 - 0.82 µm (Figure S12). For the same laser power, however, with increased scan speed of 1000 mm/s, median $|\mathbf{R}|$ increases (Figure 6b) and hence the median cell spacing prediction from the 3D coupled simulations is reduced to $\sim$ 0.52 µm (Figure 6e) with $25^{th}$ to $75^{th}$ quartiles between 0.39 - 0.61 µm (Figure S12). Evidently, the reduction in the median cell spacing and the the distribution ranges obtained from the coupled simulations are in close agreement with the ex situ SEM cell spacing distribution in Figure 6d. These quantitative results highlight the applicability of coupling high-speed IR imaging with 3D multiphysics simulations to gain valuable transient sub-surface solidification conditions. We observe a similar reduction in cell spacing with increase in laser scan speed for a laser power of 300 W (Figure S11).

Typically, in literature, results from thermal simulations are often compared with cross-sections from an experimental melt track, as performed here. This cross-section is typically extracted from the presumed 'steady-state' region, where the thermal conditions are assumed to be relatively constant with respect to the scan direction. However, this comparison neglects the possible variability in thermal conditions with respect to the scan direction. Due to the high temporal resolution of the measured surface temperatures, in combination with a high fidelity simulation approach, temporal variations in thermal conditions in three dimensions can be quantitatively assessed here. The variations are expressed as standard deviations in thermal gradient, both in magnitude and in the z-direction (towards the surface), and the predicted local cell orientation, in Figure 7.

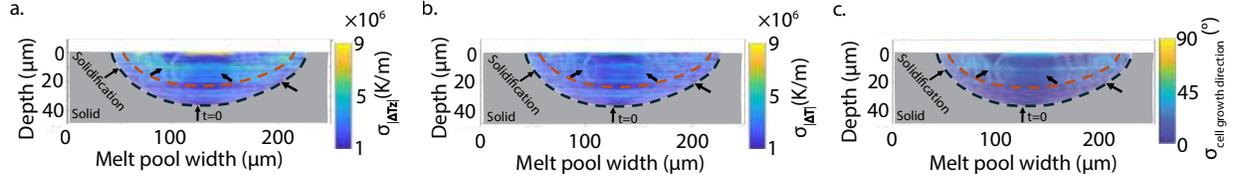

**Figure 7**: **Temporal variations in sub surface thermal conditions during 5 consecutive time steps of laser melting at 200 W and 500 mm/s.** Standard deviation of: a. thermal gradient in z-direction, b. total thermal gradient, c. predicted cell orientation. The black dashed curves represent S-L boundary at the beginning of solidification, whereas the orange dashed curve is the S-L boundary after some melt pool solidification has occurred.

The standard deviations are all at their lowest respective magnitudes at the begining of the solidification (region close to black dashed curve in Figure 7a and b) and generally increase as the solidification front progresses. Towards the tail of solidification, the standard deviation increases to $8\times10^6$ K/m (Figure 7a) for $|\nabla T_Z|$ and $5\times10^6$ K/m (Figure 7b) for $|\nabla T|$. The markedly higher variations of thermal gradients, towards the tail of melt pool, could be responsible for promoting a columnar-to-equiaxed transition (CET) or other similar microstructural change at the center of the solidified melt pool. In that regard, significant deviations in the predicted cell orientation are shown in this same region (inside the red dashed curve in Figure 7c). Overall, in the bulk of the melt pool, these variations are sufficiently low, which suggests microstructural features like cell spacings and epitaxial growth would be consistent across multiple cross-sections. Variations in thermal conditions only become sufficiently large at the tail of the melt pool. This is a region that is susceptible to microstructural transitions, such as the CET, or the formation of defects, like solidification cracking. Notably, this is also the same region of the melt pool where convective heat transfer dominates over conduction, and large variation in solidification velocity magnitudes and vectors are observed.

## Discussion

All of the foregoing observations and analyses not only validate the applicability and usefulness of the coupled modeling approach, but also illuminate new insights on transient temperature distributions within the melt pool. The predominance of convective heat transfer near the surface (Figure 2) are uniquely captured by the surface IR measurements. In these same regions, the solidification velocity rapidly increases in magnitude and standard deviation (Figure 3f), and the direction normal to the S-L interface becomes more variable (Figure 4). Large temporal variations in thermal gradients are also observed in this location (Figure 7). Significant variability of solidification conditions near the melt pool surface can ultimately manifest in a high susceptibility to microstructural heterogeneities.

We propose that the coupled modeling approach can provide an alternative path to estimate parameters for and validate microstructural models. For instance, the material-dependent constant *A* necessary to estimate cell spacings is often assumed in analytical models. In this work, a singular fit to experimental data in one melt track is applicable across all thermal conditions evaluated, and accurately predicts spatiotemporal changes in cell sizes. Though not explicitly conducted here, this approach would rapidly substantiate microstructure prediction maps, such as the CET. This approach can be extended to new materials that are still missing measurements of key properties while retaining high accuracy. Having elaborated this possibility, we further emphasize that methodology can be applied to other melting scenarios, such as in-situ monitoring of additive manufacturing with powder feedstock. Some notable differences between powder feedstock, and the wrought substrate used here, would be the thermophysical properties of the solid powder, and laser absorption impacted by the powder morphology. In both cases, the modeling approach directly accounts for these changes. Absorption can be inferred through the heat flux determined from experimental IR measurements (Figure 2), and the thermal conductivity of the liquid remains identical within the melt pool. The parametric investigation on the impact of thermophysical properties (Figure 1) shows that the coupled modeling approach should work well even with uncertainties in thermophysical properties, such as with powder.

In summary, we report a coupled method to generate three-dimensional temperature distributions that are directly validated by in-situ IR with high spatio-temporal resolution. Complexity and uncertainty typically associated with this type of computation is reduced by the use of IR measurements as the boundary condition. With sufficiently high spatio-temporal resolution of IR imaging, the coupled approach can quantify the local spatial and temporal variations in the 3D melt pool profile and solidification conditions (thermal gradients and solidification velocities). This methodology is exemplified by evaluating laser track melting experiments, relevant to additive manufacturing conditions, on a Ni superalloy. Through coupled microstructural analyses, material-dependent constants for predictive microstructure maps are determined. The multimodal three-dimensional data elucidate the role of convective heat transfer in the spatial and temporal variability of thermal conditions. The approach can be universally extended to different manufacturing processes and next generation materials, inspiring the possibility for microstructure prediction and control in all scenarios.

# Acknowledgments

Authors thank Wes Autran, Alex Côté, Vince Morton, Andrew Niehaus, and Hector Moreno from Telops for discussions and suggestions on the temperature conversion process. Disclaimer: This report was prepared as an account of work sponsored by an agency of the United States Government. Neither the United States Government nor any agency thereof, nor any of their employees, makes any warranty, express or implied, or assumes any legal liability or responsibility for the accuracy, completeness, or usefulness of any information, apparatus, product, or process disclosed, or represents that its use would not infringe privately owned rights. Reference herein to any specific commercial product, process, or service by trade name, trademark, manufacturer, or otherwise does not necessarily constitute or imply its endorsement, recommendation, or favoring by the United States Government or any agency thereof. The views and opinions of authors expressed herein do not necessarily state or reflect those of the United States Government or any agency thereof.

**Funding:** The IR imaging part of this work is supported by the National Science Foundation through the Materials Research Science and Engineering Center (MRSEC) at UC Santa Barbara: NSF DMR-2308708. The conversion of IR-imaging to COMSOL simulations and analysis was supported by the Office of Naval Research grant number N00014-24-1-2086. The laser melting experimental setup and materials characterization are based upon work supported by the Department of Energy, National Nuclear Security Administration under Award Number(s) DE-NA0004152. KM was supported by a National Science Foundation (NSF) Graduate Research Fellowship under Grant No. 2139319. The research reported here made use of the shared facilities of the Materials Research Science and Engineering Center (MRSEC) at UC Santa Barbara: NSF DMR–2308708. The UC Santa Barbara MRSEC is a member of the Materials Research Facilities Network (www.mrfn.org).

**Author contributions:**   Y.Z. and V.K. conceptualized the idea with preliminary work. V.K., K.M.M., and H.P. designed and carried out the experiments. K.M.M performed SEM and EBSD characterization. V.K. performed IR imaging data interpretation and COMSOL Multiphysics simulations. M.G, A.B, H.P, Y.Z, K.M.M, T.M.P. contributed to data analysis. V.K. wrote original draft and K.M.M, Y.Z., T.M.P, V.K. contributed to manuscript revision. T.M.P and Y.Z. acquired funding and supervised this work.

**Competing interests:**   There are no competing interests to declare.

**Data and materials availability:**   All data needed to evaluate the conclusions in the paper are presented in the paper and/or Supplementary Materials.

# Supplementary materials

The pdf file includes

Note 1 to 6

Figures S1 to S14

Movie S1

Tables S1 and S2